\documentclass[sigplan]{acmart}

\acmConference[EMC2'19]{Workshop on Energy Efficient Machine Learning and Cognitive Computing for Embedded Applications}{February 17, 2019}{Washington D.C., USA}
\acmYear{2019}
\acmISBN{} 
\acmDOI{} 
\startPage{1}

\setcopyright{none}

\usepackage{booktabs}   
\usepackage{subcaption} 
\usepackage{multirow}

\usepackage{url}

\usepackage{breakurl}
\usepackage{hyperref}

\def\fixme#1{\typeout{FIXED in page \thepage : {#1}}
\bgroup \color{red}{[FIXME: {#1}]} \egroup}
\begin{document}

\title[Integrating NVDLA with RISC-V SoC on FireSim]{Integrating NVIDIA Deep Learning Accelerator (NVDLA) with RISC-V SoC on FireSim}         



\author{Farzad Farshchi}
\affiliation{
  \institution{University of Kansas}
}
\email{farshchi@ku.edu}

\author{Qijing Huang}
\affiliation{
  \institution{University of California, Berkeley}
}
\email{qijing.huang@berkeley.edu}

\author{Heechul Yun}
\affiliation{
  \institution{University of Kansas}
}
\email{heechul.yun@ku.edu}

\begin{abstract}
NVDLA is an open-source deep neural network (DNN) accelerator which has received a lot of attention by the community since its introduction by Nvidia. It is a full-featured hardware IP and can serve as a good reference for conducting research and development of SoCs with integrated accelerators. However, an expensive FPGA board is required to do experiments with this IP in a real SoC. Moreover, since NVDLA is clocked at a lower frequency on an FPGA, it would be hard to do accurate performance analysis with such a setup. To overcome these limitations, we integrate NVDLA into a real RISC-V SoC on the Amazon cloud FPGA using FireSim, a cycle-exact FPGA-accelerated simulator. We then evaluate the performance of NVDLA by running YOLOv3 object-detection algorithm. Our results show that NVDLA can sustain 7.5 fps when running YOLOv3. We further analyze the performance by showing that sharing the last-level cache with NVDLA can result in up to 1.56x speedup. We then identify that sharing the memory system with the accelerator can result in unpredictable execution time for the real-time tasks running on this platform. We believe this is an important issue that must be addressed in order for on-chip DNN accelerators to be incorporated in real-time embedded systems.
\end{abstract}

\keywords{NVDLA, FireSim, RISC-V, FPGA, Cloud, DNN, Embedded Systems}  

\maketitle

\section{Introduction}
In recent years, deep neural networks (DNNs) are increasingly used in sophisticated embedded/robotics systems, such as self-driving cars and drones, due to their superior performance in solving complex perception and control tasks over traditional methods. However, DNNs are computationally intensive and have large memory requirement \cite{gao2017tetris}, while embedded systems generally impose strict constraints on power, size, and weight of their computing platforms.


To improve performance and efficiency of DNN processing, especially inference operations, in embedded systems, hardware DNN accelerators are being incorporated into embedded system-on-chips (SoCs). 
Recently, Nvidia introduced an open-source inference engine, called Nvidia Deep Learning Accelerator (NVDLA) \cite{nvdla}, and integrated it into their Xavier SoC platform. They also announced a joint plan with Arm corporation to integrate NVDLA blocks in future ARM SoCs~\cite{nvdla_arm}. 
In August 2018, a semiconductor startup SiFive announced that they have integrated the NVDLA block into their open-source SoC platform \cite{nvdla_sifive}. In the demo presented at Hot Chips'18, NVDLA is implemented on an FPGA, which is connected to  SiFive's Freedom U540 SoC platform\cite{fu540} via an off-chip bus.\footnote{The first author performed the FPGA integration for the demo during his internship at SiFive in summer 2018.}

We believe SiFive's integration of NVDLA is especially a useful platform for conducting research thanks to its open-source nature. However, the platform has several limitations: First, the FPGA board used to implement this platform costs about \$7k \cite{vcu118}. This is an expensive piece of equipment for many research institutions and even some smaller companies. Second, a design implemented on FPGA has to be clocked at a lower frequency comparing to the equivalent ASIC implantation, however, DRAMs in both implementations are clocked at about the same frequency. Thus, the design implemented on FPGA sees relatively faster DRAM. Third, the current open-source release of SiFive's SoC platform does not support an L2 cache. In particular, the second and third limitations make performance-related research difficult on the platform.

To solve the aforementioned limitations, in this paper, we have integrated NVDLA into FireSim \cite{karandikar2018firesim}. FireSim is an FPGA-accelerated full-system simulator, which runs on the Amazon cloud FPGAs. FireSim's target design is derived from the open-source RISC-V-based Rocket Chip SoC~\cite{Asanović:EECS-2016-17}. In FireSim, a special transform is applied on the target design that decouples it from the DRAM controller of the host FPGA and adds a memory model to the simulation environment. This allows us to model a realistic DRAM subsystem. We also model a last-level cache (LLC) similarly. Our FireSim integration solves the second and third limitations of SiFive's real FPGA-based integration. Moreover, since FireSim runs on the cloud, a research group may pay based on the hourly usage, without needing to purchase an expensive FPGA board. To compare with the cost of an FPGA board, the rate for on-demand access to the smallest Amazon FPGA instance is \$1.65 per hour. Our integration of NVDLA with RISC-V SoC on FireSim is publicly available as open-source on GitHub.\footnote{\url{https://github.com/CSL-KU/firesim-nvdla}}

The rest of the paper is organized as follows. Section \ref{sec:back} describes the background on NVDLA and FireSim. Section \ref{sec:int} describes the NVDLA integration. In Section \ref{sec:perf}, we used NVDLA in FireSim to demonstrate how this platform can be used for performance analysis. We particularly focused on analyzing the performance achieved by sharing the LLC with NVDLA and interference caused by sharing the memory system between the CPU cores and the NVDLA.

\section{Background} \label{sec:back}

In this section, we provide a brief background on NVDLA and FireSim. 

\subsection{NVDLA}
\begin{figure}[h]
\centerline{\includegraphics[width=0.8\linewidth]{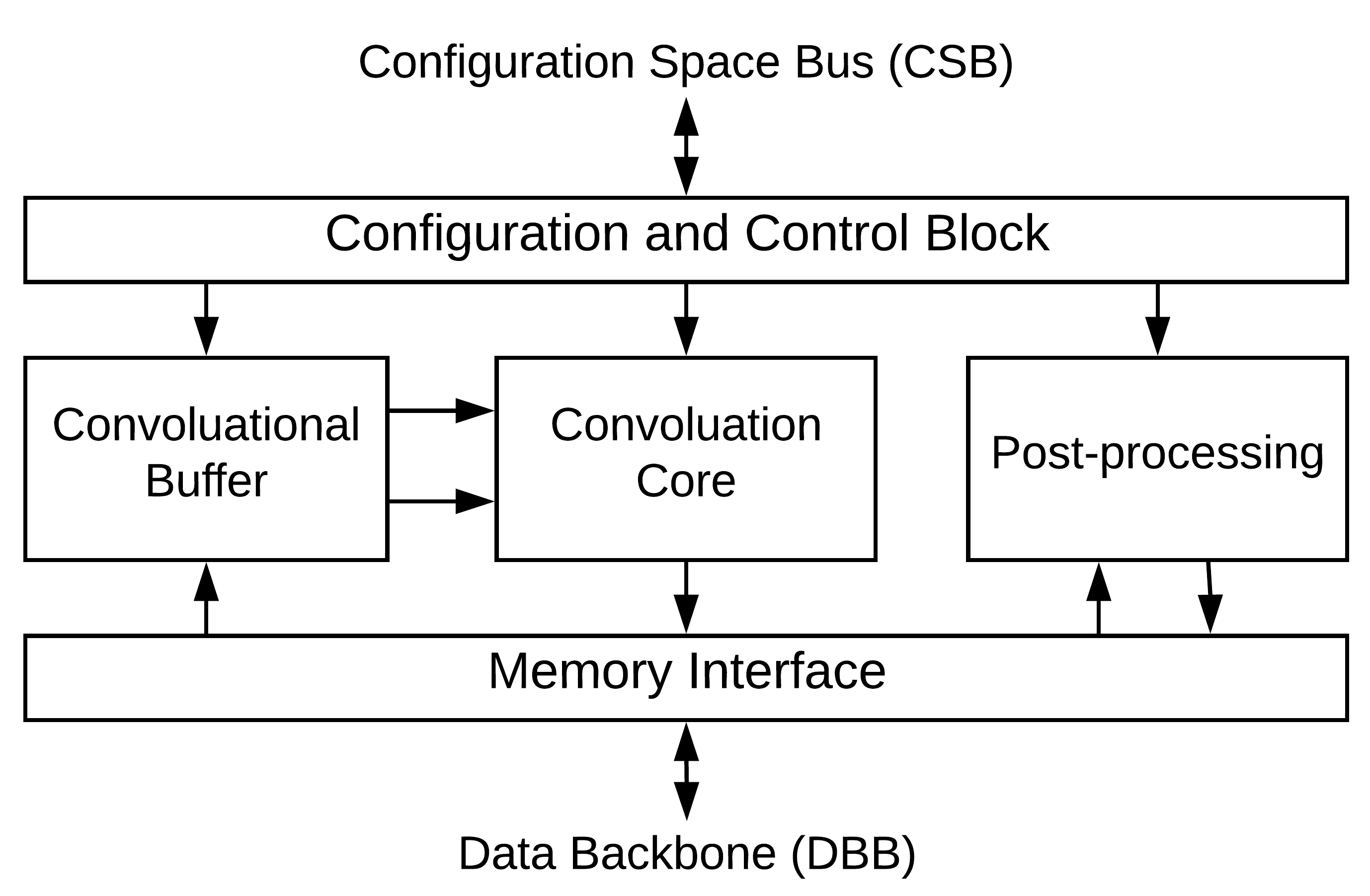}}
\caption{NVDLA architecture. Adopted from \cite{nvdla_slide}.}
\label{fig:nvdla}
\end{figure}

NVDLA is an industry-grade open-source DNN inference engine, developed by Nvidia\cite{nvdla}. It is released as Verilog source code and  is configurable at the build time to meet different performance, power, and  area  trade-offs. NVDLA mainly targets embedded systems and IoT devices with limited power budget. Figure~\ref{fig:nvdla} shows a high-level architecture of NVDLA. As it can be seen in this figure, NVDLA has three major top-level blocks. Convolutional Core is consisted of multiply-accumulate (MAC) units for matrix-matrix multiplication in convolutional and fully-connected layers of a DNN. The input activations and filter weights are stored in Convolutional Buffer which are then fed into Convolutional Core. Post-processing unit is comprised of sub-units which perform various processes such as pooling and applying non-linear activation functions. These three blocks are programmed and controlled by Configuration and Control Block, which is accessed by the host processor via Configuration and Space Bus (CSB) interface. All the processing units are connected to Memory Interface block. This block arbitrates the access to main memory via  Data Backbone (DBB) interface. The CSB and DBB interfaces are further described in Section~\ref{sec:int}.

\subsection{FireSim}
FireSim is a fast cycle-exact system simulator which runs on cloud FPGA. In FireSim, the simulated hardware is derived from the actual RTL of the design, thus, the same RTL can be pushed through a VLSI flow to fabricate the chip, and since the simulation is running on FPGA, it is orders of magnitude faster than software-based simulation. In addition, FireSim can be used for debug and trace generation. Lastly, since it is running on the cloud, there is no need to pay the upfront cost to buy FPGA boards. These traits make FireSim more preferable in many aspects comparing to software-based simulators to be used in areas of computer architecture and system research. For instance, it provides the capabilities of faster and more accurate performance analysis and true hardware/software co-design.
\section{NVDLA Integration} \label{sec:int}

For this integration, we choose the \emph{nv\_large} configuration of NVDLA which has 2048 MAC units and a 512 KiB convolutional buffer. NVDLA connects to the rest of the SoC via these interfaces:

\textit{Configuration Space Bus (CSB)}: A low-bandwidth slave interface which provides access for the host processor to configure NVDLA and read the status. The CSB interface implements a simple custom protocol. An adapter is supplied in the NVDLA code repository to convert this simple protocol to ARM Advanced Peripheral Bus (APB)~\cite{apb}.

\textit{Data Backbone (DBB)}: A high-bandwidth master interface which is connected to the memory system. NVDLA uses this interface to access the main memory. DBB implements ARM Advanced eXtensible Interface (AXI)~\cite{axi} protocol.

\textit{IRQ}: A 1-bit interrupt line which is asserted when NVDLA finishes an on-going task or an error occurs. 

\begin{figure}[h]
\centerline{\includegraphics[width=0.8\linewidth]{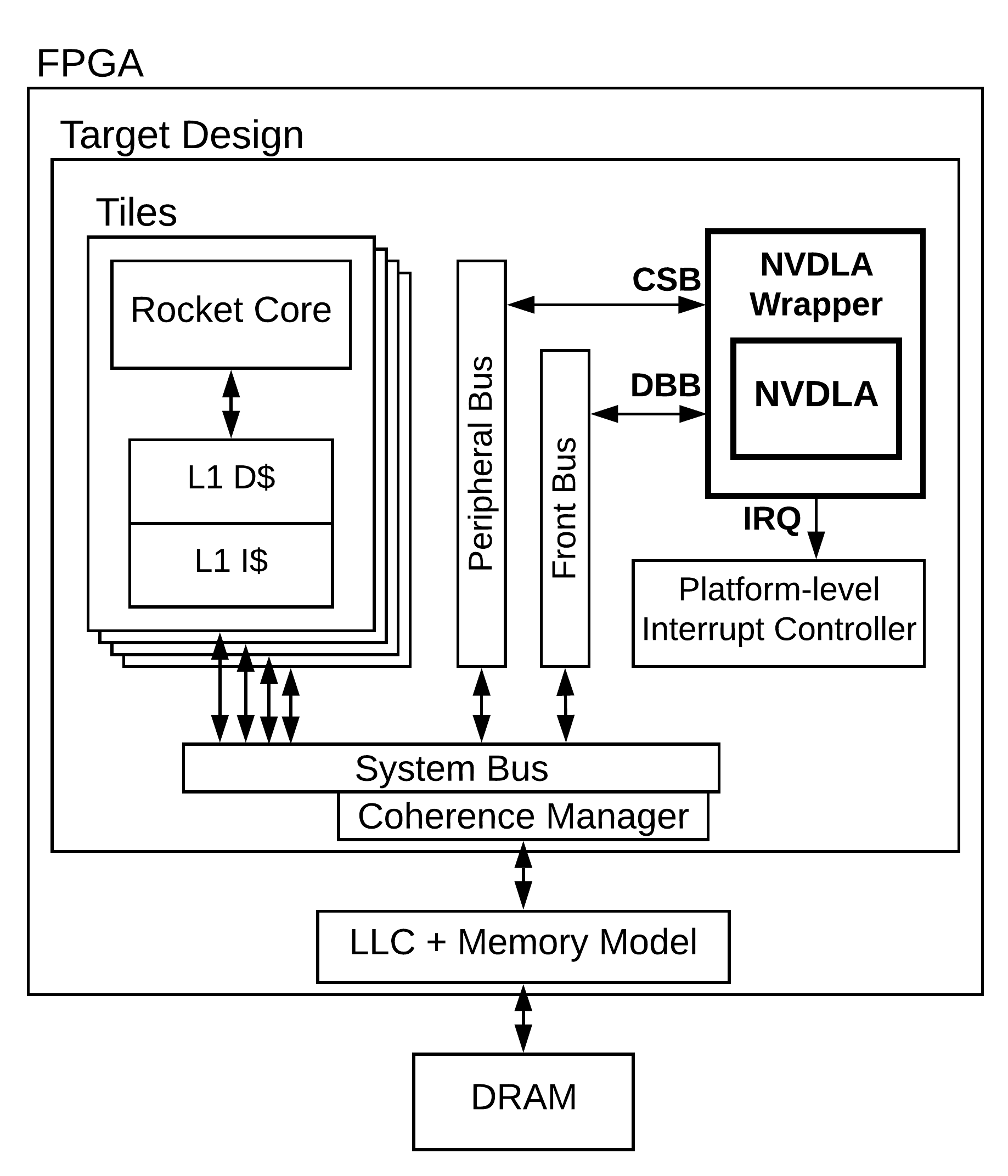}}
\caption{Simulator overview with our NVDLA integration.}
\label{fig:system}
\end{figure} 

FireSim's target design is derived from Rocket Chip SoC generator~\cite{Asanović:EECS-2016-17}, which is written in Chisel hardware construction language~\cite{bachrach2012chisel} and uses TileLink bus protocol \cite{tilelink} for on-chip communication.
Figure~\ref{fig:system} shows the overview of the simulator with our NVDLA integration. We describe the blocks in this figure which are important for understanding our integration:

\textit{Front Bus}: A TileLink switch that arbitrates non-CPU masters and connects them to the memory system. NVDLA is the only non-CPU master in our design, therefore, Front Bus is only connected to DBB interface of NVDLA.

\textit{Peripheral Bus}: A TileLink switch that connects the slave devices and maps them to a memory region. We connect CSB interface to Peripheral Bus in order for the processor to access the memory-mapped registers of NVDLA. To save space, the other peripherals connected to this switch are not shown in Figure~\ref{fig:system}.

\textit{Platform-level Interrupt Controller (PLIC)}: A prioritized interrupt controller which routes the interrupt sources of devices to CPU cores. We connect interrupt line of NVDLA and other devices (not shown in Figure~\ref{fig:system}) to PLIC.

\textit{NVDLA Wrapper}: Since Rocket Chip uses TileLink bus protocol for on-chip communication, we create this wrapper layer that converts bus transactions between the NVDLA (which uses ARM APB and AXI bus protocols) and the Rocket Chip SoC.

\textit{LLC and Memory Model}: This model is not part of the Rocket Chip SoC and is added to the simulation environment by FireSim to model accurate LLC and main memory timing. Note that this model is written in Chisel and is implemented on FPGA to avoid being a bottleneck for the simulation speed. The FireSim user can choose the exact behavior from a variety of models such as an ideal memory with constant access latency or a DRAM model. The LLC model is configurable at the runtime. The number of sets, ways, and block size can be modified without having to rebuild the FPGA image. We use this feature to analyze the performance of NVDLA for multiple different number of sets and block sizes in Subsection~\ref{sec:cacheperf}.

Thanks to the fast integration and development capabilities that Chisel offers, we find it convenient to reuse the SiFive's NVDLA integration code for our purpose. We still need to create the NVDLA simulation model and modify the FireSim and Amazon Web Services (AWS) FPGA flow to integrate NVDLA on FireSim. We describe our method of creating the NVDLA simulation model in the following.

\subsection{Creating the NVDLA Simulation Model}
FireSim uses FAME-1 transform~\cite{tan2010case, kim2016strober} to translate the target RTL to the target model and create a token-based simulator. In each target cycle, the target model reads a token on its input and generates a token on the output. However, there can be cycles in which a token is not available to be consumed by the target model. For instance, the memory model can be configured to service the memory accesses in 10 cycles while it may take longer for the FPGA DRAM to actually read the data in order for the memory model to pass it to the target model. In this case, FireSim stalls the target model until the FPGA DRAM reads the data and a token is available to be consumed by the target model.

\begin{figure}[h]
    \centering
    \begin{subfigure}[b]{0.4\textwidth}
        \includegraphics[width=\textwidth]{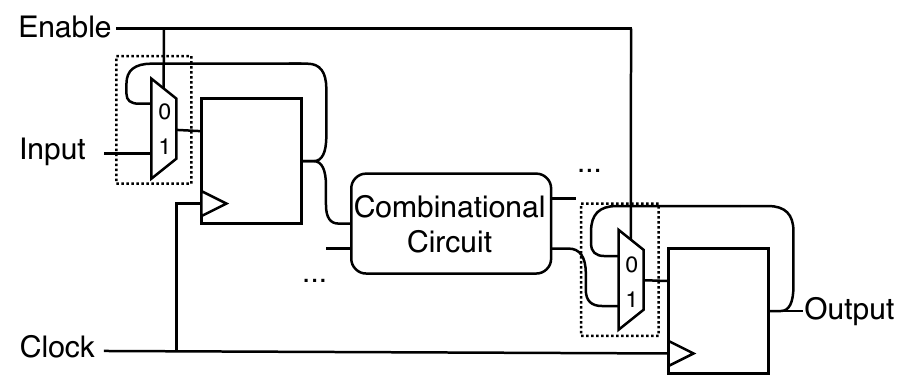}
        \caption{Register enable.}
        \label{fig:registerenable}
    \end{subfigure}
    
    \vspace{1em}
        
    \begin{subfigure}[b]{0.4\textwidth}
        \includegraphics[width=\textwidth]{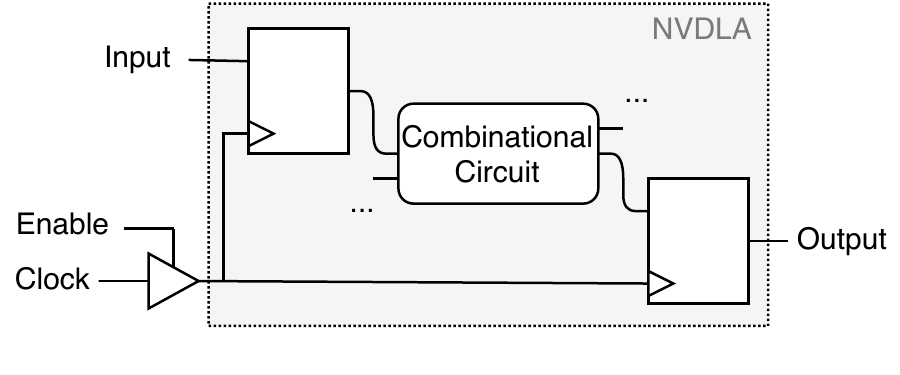}
        \caption{Clock-gating.}
        \label{fig:bufgce}
    \end{subfigure}
    \caption{FAME-1 transform with (a) register enable (b) clock-gating.} 
\end{figure}

In order to stall the target model, FireSim runs a pass in the back-end of Chisel compiler, which adds a global enable signal and an input mux for each register in the target design as shown in Figure~\ref{fig:registerenable}. By deasserting the enable signal, the target model is stalled at the cycles which a token is not available on the input. Since NVDLA source is written in Verilog, the pass cannot be directly applied. Instead, we add a new pass in the Chisel back-end to identify the Verilog modules in the design and apply the FAME-1 transform by clock-gating as shown in Figure~\ref{fig:bufgce}. This new pass identifies the input clock source of the NVDLA Verilog module and adds a clock buffer with an enable signal. Once the enable signal is deasserted, the clock input remains low to stall the NVDLA model.

\section{Performance Analysis} \label{sec:perf}
In this section, we demonstrate the capabilities of our integrated platform for performance analysis. Specifically, we analyze the effect of interference between the NVDLA and RISC-V CPU cores in accessing the shared memory hierarchy including LLC and DRAM.

For this purpose, we choose the state-of-the-art YOLOv3 object detection algorithm \cite{redmon2018yolov3} as a benchmark. YOLOv3 incorporates a DNN that needs 66 billion operations to process a $416\times416$ frame. Table~\ref{tbl:platconf} shows our platform configuration parameters. Note that NVDLA block is clocked at the same frequency with the processor, due to a limitation of current FireSim implementation, which requires all hardware blocks to have synchronous clock sources. In a real chip, NVDLA may be operated at a lower frequency. We hope that this will be fixed in near future as FireSim is planned to support having asynchronous clocks.

\begin{table} [h]
  \centering
  \caption{Baseline platform configuration.}
  \begin{tabular}{|c|c|}
    \hline
    Processor 	& Quad-core, in-order, single-issue, 3.2 GHz \\
    \hline
    NVDLA       & 2048 INT8 MACs, 512 KiB buffer, 3.2 GHz \\
    \hline
    L1 I/D\$ & Private 16/16 KiB, 4-way, 64 B block \\
    \hline
    LLC & Shared 2 MiB, 8-way, 64 B block \\
    \hline
    DRAM & 16 GiB DDR3, 4 ranks, 8 banks, FR-FCFS \\
    \hline
  \end{tabular}
  \label{tbl:platconf}
\end{table}

On this configuration, we first measure the baseline performance of NVDLA.
The time that it takes to process a frame on this platform is 133 ms out of which 67 ms is spent on NVDLA and the rest on the processor. Some of the computations which are not supported by NVDLA are executed by the processor. These are particularly upsampling, floating-point to integer conversion (and vice-versa), and custom YOLO layers. We optimized these layers with OpenMP to utilize all four processor cores. 

To calculate the speedup achieved by using NVDLA, we measure the performance of YOLOv3 when Rocket Cores are performing all the computations. In addition, we run the benchmark on two other platforms: 1) Intel Xeon E5-2658 v3 CPU (2 sockets; 24 cores/48 threads in total) and 2) Nvidia Titan Xp GPU. The benchmark is running multithreaded on Rocket Core and Xeon platforms. Note that computations are performed with 8-bit integer precision on NVDLA and with single-precision floating-point on the rest of the platforms. We use the open-source Darknet~\cite{darknet13} neural network framework to conduct our experiments.  

\begin{figure}[h]
\centerline{\includegraphics[width=0.8\linewidth]{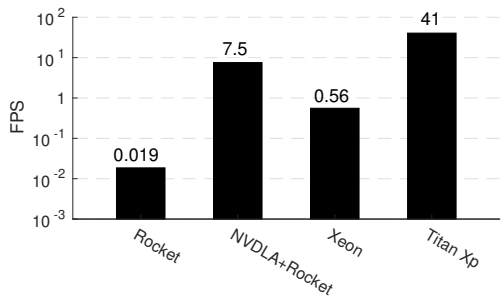}}
\caption{The performance of YOLOv3 object-detection algorithm on different platforms.}
\label{fig:yolo-fps}
\end{figure}

Figure~\ref{fig:yolo-fps} shows the number of frames per second (FPS) on each platform. It can be seen that using NVDLA, the performance is improved by 407x comparing to the case, which Rocket Cores are performing all the computations. In this experiment, Titan Xp can extract the highest level of parallelism and achieves the FPS of 41 thanks to its 3840 CUDA cores. This is 5.5 times faster than NVDLA, meanwhile, according to power and area numbers published on \cite{nvdla} and \cite{titan}, NVDLA consumes much less power and chip area than Titan Xp and, therefore, can provide the capability of running computationally-intensive tasks in real-time on low-cost platforms with limited power and size. 

\subsection{Effect of Last-level Cache on Performance} \label{sec:cacheperf}
Sharing the LLC with the DNN accelerator has some benefits comparing to using a scratch pad. First, chip area can be saved by removing a large SRAM which implements the scratch pad. Second, less programming effort is needed as the data movement is managed by the cache controller. In this experiment, we examine the performance improvement achieved by sharing the LLC. We vary the LLC size and measure NVDLA speedup with respect to a design with no LLC. The LLC size is changed by varying the number of cache sets. For each LLC size, we also varied the size of cache blocks (32B, 64B and 128B) to see their impact to the speedup. The rest of the parameters are the same as the baseline configuration in Table~\ref{tbl:platconf}. Since our focus is on the performance of NVDLA, the speedup is measured for the layers of YOLOv3 that run on NVDLA.

\begin{figure}[h]
\centerline{\includegraphics[width=0.8\linewidth]{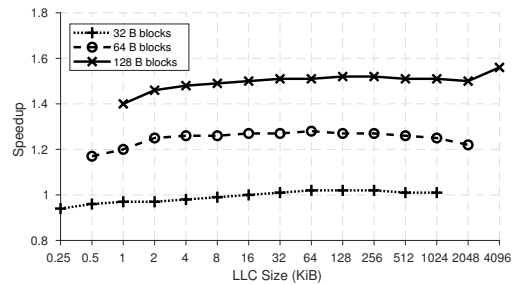}}
\caption{NVDLA speedup achieved by using the last-level cache.}
\label{fig:llc-speed}
\end{figure}

Figure~\ref{fig:llc-speed} shows the results. First, let us focus on the baseline '64B blocks' results. For 64-byte cache blocks configuration, adding an LLC considerably improves performance (up to 1.28x speedup). Interestingly, however, the size of the LLC does not have much differences. Specifically, when the LLC size is at 0.5 KiB, it achieves 1.17x speedup, which is not much different from the maximum speedup of 1.28x, which is achieved at the LLC size of 64 KiB.

As shown above the performance of NVDLA is not very sensitive to the LLC size. This is due to the large convolutional buffer that captures most of the temporal locality in NVDLA memory accesses. However, the results in Figure~\ref{fig:llc-speed} shows that NVDLA performance varies to a much larger extent when the cache block size is varied. For instance, the speedup achieved by using LLCs with the constant capacity of 1024 KiB and different block sizes of 32B, 64B, and 128B is 1.01x, 1.25, and 1.51x respectively. The maximum speedup which can be achieved by using the LLC is 1.56x for a 4096 KiB cache with 128-byte blocks. The reason is that the minimum burst size of NVLDA memory accesses is 32 bytes, therefore, having cache blocks larger than 32 bytes helps to reduce the latency for larger bursts or later accesses to nearby memory locations. This shows most of the benefit of sharing the LLC comes from capturing the spatial locality in NVDLA memory references. Therefore, it is likely that hardware prefetching further improves NVDLA performance on this platform as it can bring useful data into the cache before it is accessed by NVDLA.



\subsection{Effect of Shared Memory Interference}
Since NVDLA and the CPU are sharing the memory system, it is likely that they  interfere with one another when accessing the memory. This in turn can result in unpredictable latency in execution of the tasks running on NVDLA.
This problem becomes more important on real-time systems in which the total correctness of operation depends on both the logical correctness of computation and the time that it is performed.

We use Bandwidth Write (BwWrite) benchmark~\cite{valsan2016taming} to study the interference caused by the tasks running on the processor. BwWrite is a synthetic benchmark which writes data to sequential memory addresses to generate the maximum memory traffic by the processor. Is it possible to set the working set size (WSS) of BwWrite to target different levels of the memory hierarchy. We co-schedule BwWrite with YOLOv3 running on NVDLA and vary the set size of BwWrite to either fit into L1 cache, LLC, or DRAM. The number of co-scheduled BwWrites is also varied from 1 to 4.
We pinned BwWrites to specific cores by setting the CPU affinity in Linux.
The execution time on NVDLA is measured and normalized to \emph{solo} execution time i.e. NVDLA running in isolation with no BwWrite running on the cores.

\begin{figure}[h]
\centerline{\includegraphics[width=0.8\linewidth]{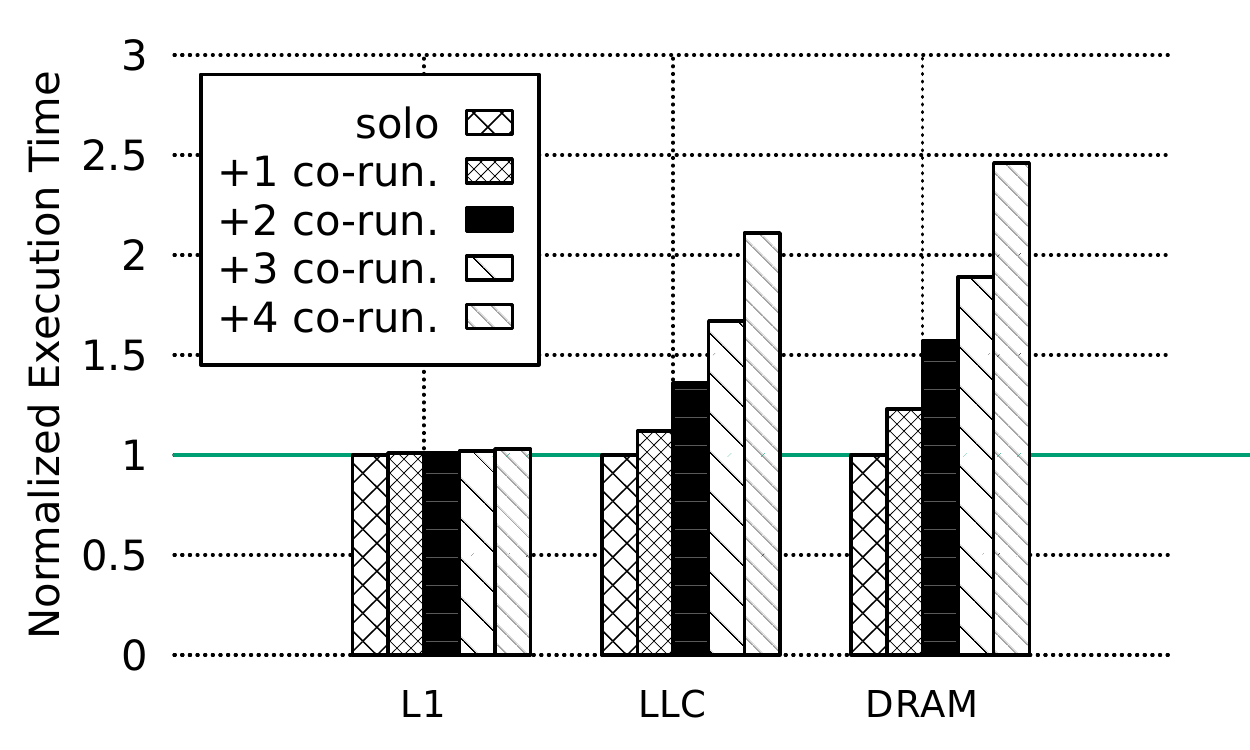}}
\caption{Normalized execution time of NVDLA. The x-axis denotes the working-set size of BwWrite co-runners.}
\label{fig:bw-effect}
\end{figure}

Figure~\ref{fig:bw-effect} shows that there is no slowdown when the working set of BwWrite fits into the L1 cache. The reason is each core has its own private data cache and as long as the WSS is smaller than L1, there is no access to the shared memory. When the WSS of BwWrite is LLC-fitting and is larger than L1 cache size (denoted with \emph{LLC} in the figure), it takes longer for NVDLA to finish the job as the memory accesses are delayed due the concurrent accesses by the cores to the shared bus and the LLC. This can results in up to 2.1x slowdown when four BwWrites are running in parallel. The slowdown increases to 2.5x for 4 co-runners when WSS is DRAM-fitting since interference in the DRAM controller scheduler queue and the DRAM banks adds more delay to NVDLA memory access time.

\section{Conclusion}
In this paper, we described our integration of NVDLA into a RISC-V multicore SoC on FireSim. The integrated SoC platform runs on Amazon cloud and does not require any physical FPGA board. 
Furthermore, thanks to FireSim, the performance critical shared cache and memory subsystems are easily configurable. Therefore, we believe that our platform is a flexible and cost-effective solution for conducting research. 
As a case study, we ported YOLOv3 and evaluated its acceleration performance, compared to GPU and CPU solutions, and under various cache and memory configurations. We find that (1) NVDLA provides good acceleration performance, especially considering its low power consumption, (2) larger cache block size and/or hardware prefetcher is desirable for performance, and (3) the impact of shared memory interference between CPU and NVDLA is significant---which can be especially problematic for critical real-time embedded systems found in automotive applications---suggesting the need of additional QoS mechanisms (e.g., \cite{farshchi2018deterministic,iyer2007qos, grot2012qos}).


\begin{acks}                            
This research is supported in part by NSF grants CNS 1718880, CNS
1815959, and NSA Science of Security initiative contract no. \#H98230-18-D-0009.
\end{acks}

\bibliography{references}



\end{document}